\documentclass[runningheads]{llncs}
\usepackage{graphicx}
\begin{document}
\title{LACLICHEV: Exploring the History of Climate Change in Latin America within Newspapers Digital Collections\thanks{We thank the master student Santiago Ruiz Angulo of the Universidad Aut\'onoma de Guadalajara who implemented the first version of LACLICHEV during his internship at the Barcelona Super Computer Centre, Spain funded by the CONACYT “beca mixta” fellowship program of the Mexican government.}}
\author{Genoveva Vargas-Solar\inst{1} \and
José-Luis Zechinelli-Martini \inst{2}   \and 
Javier A. Espinosa-Oviedo\inst{3} \and
Luis M. Vilches-Blázquez \inst{4}}
\authorrunning{G. Vargas-Solar et al.}
\institute{CNRS, LIRIS 
Campus de la Doua, 69622 Villeurbanne, France \\ 
\email{genoveva.vargas-solar@liris.cnrs.fr}
\and
Fundación Universidad de las Américas Puebla, 72820 San Andrés Cholula, Mexico\\
\email{joseluis.zechinelli@udlap.mx}\\
\and
Université Lumière Lyon 2, ERIC Lab, Lyon, France\\
\email{javier.espinosa-oviedo@univ-lyon2.fr}\\
\and
Centro de Investigación en Computación, IPN, 07738 Ciudad de México, Mexico\\
\email{lmvilches.blazquez@gmail.com}}

\maketitle              % typeset the header of the contribution
\begin{abstract}
This paper introduces LACLICHEV (Latin American Climate Change Evolution platform ), a data collections exploration environment for exploring historical newspapers searching for articles reporting meteorological events. LACLICHEV is based on data collections' exploration techniques combined with information retrieval, data analytics, and geographic querying and visualization. This environment provides tools for curating, exploring and analyzing historical newspapers articles, their description and location, and the vocabularies used for referring to meteorological events. The objective being to understand the content of newspapers and identifying possible patterns and models that can build a view of the history of climate change in the Latin American region. 
\vspace{-0,3cm}
\end{abstract}
%
%
%

%----------------------------------------------------------------------------
\section{Introduction}
%----------------------------------------------------------------------------
% Ninety-seven per cent of climate scientists agree that climate-warming trends over the past century are very likely due to human activities (NASA, 2018a). Some observations report and studies reveal that the planet's average surface temperature has risen about 2.0 degrees Fahrenheit (1.1 degrees Celsius) since the late 19th century. The hypothesis is that this change has been mainly driven by increased carbon dioxide and other human-made emissions into the atmosphere. 

Historical analysis of climate behaviour can provide  conclusions about climatologic phenomena and Earth climate behaviour. There exist several scientific efforts to study the history of climate change. The Climate of the Past \cite{EUsciences}, for example, is an international scientific journal dedicated to the publication and discussion of research articles, short communications, and review papers on Earth's climate history. The journal covers all temporal scales of climate change and variability, from geological time through to multidecade studies of the last century. The Government of Canada provides access to historical observations on climate in Canada starting from 1840 \cite{Canada} . However, these data collections are disconnected and use different reference variables and observation criteria. Even if there is an increasing interest in analysing digital data collections for performing historical studies on climatologic events, the history of climate behaviour is still an open issue that has not revealed missing knowledge.

Technological advances have enabled the understanding of phenomena and complex systems by collecting many different types of information.
Data collections are exported under different releases with different sizes, formats (e.g., CSV, text, excel), sometimes with various quality features. Tools helping to understand, consolidate and correlate data collections are crucial.
%
% Long historical data studies could make it possible to compute more complete models of climatic phenomena and the conditions in which they emerged. However, meteorology is a young science that started around the 19th century. It is supported by more or less recent data, making it challenging to run an analysis that can give more historical pictures of climatic evolution and its implications using observations instead of extrapolations. Those willing to promote changes in the behaviour of society and industry for reducing emissions that have a role in climate change must convince civil society of the importance of the challenges. For this reason, our work addressed the problem of collecting and analyzing the history of meteorological events to explore how they were described, lived and perceived by civil society.
%
% In this sense, the digitalization of data collections has an increasingly vital role in collecting vast amounts of \textit{hidden} data. Thus, taking into account that digital archives become more easily accessible every time and contain explicit and implicit spatiotemporal information, researchers in GI-Science are becoming aware of these new data sources. Moreover, 
Digital data collections make it possible to have an analytic vision of the evolution of environmental, administrative, economic and social phenomena. In this context, our work deals with data collections that report the emergence of meteorological events (e.g., temperature changes, avalanches, river flow growth, volcano eruptions).
%sometimes, they relate them to their causes and consequences. 
%However, the digitized collections present some issues.
% On the one hand, they are often riddled with OCR errors that hamper the performance of information retrieval systems. Therefore, handling OCR errors is one of the two significant problems for information retrieval from collections of historical documents. On the other hand, another problem of these sources is related to historical change in languages since digitized texts are written in the language of their origin.

This paper introduces a Latin American Climate Change Evolution platform called LACLICHEV. 
The objective is to expose and study the history of climate change in Latin America. 
The hypothesis is that history (in Latin America) is contained in newspapers articles lying in digital collections available in national libraries of four countries, namely Mexico, Colombia, Ecuador and Uruguay.
LACLICHEV addresses three issues: \\
%(see Figure \ref{fig:problem}):
\noindent
(i) First, newspaper archaeology by chasing articles talking about climatological events using specific vocabulary to discover as many articles as possible. The challenge is choosing the adequate vocabulary to increase chances to get articles actually talking about climatologic events. \\
\noindent
(ii) Second, once an article talks about a climatologic event, it is Geo-Temporal tagged with metadata specifying what happened, where and when it happened, its duration and geographical extent. The objective here is to build a climatologic event history.\\
\noindent
(iii) Finally, on top of this history, the objective is to run analytics questions and visualize results in maps, given that the content is highly spatial. 

% \begin{figure}[h]
% \begin{center}
% \includegraphics[width=0.90\textwidth]{figures/laclichev-overview.jpg}
% \caption{Problem Statement}
% \label{fig:problem}
% \end{center}
% \end{figure}
%\vspace{-0.8cm}
% This one is a data collections exploration environment that applies data collections exploration techniques combined with efficient retrieval, data analytics, and visualization for understanding the content of articles that report historical meteorological events. These data with high geospatial and temporal content were aggregated and consolidated into maps giving a one-shot idea of the history of meteorological events through the empiric eyes of civil society before the emergence of meteorology as a science. 

% To track events, we decided to explore newspapers for searching articles that could report such events, the conditions in which they happened, their duration, the places in which they happened and their impact in terms of an approximate number of casualties, the kind of damages, etc. As an experimental scenario, we chose the XVIII and XIX centuries, which defines a golden age for newspapers in Latin American countries, namely, Mexico, Colombia, Ecuador and Uruguay.

The remainder of this paper is organized as follows. Section \ref{sec:architecture} introduces the curation process that we propose for historical newspapers articles potentially reporting on climatologic events.
%LACLICHEV, its general architecture and its main functions for exploring data collections. %LACLICHEV provides an API (Application Programming Interface) and the back-end infrastructure for exploring data collections through a Jupyter Notebook. 
%Section \ref{sec:model} 
%It introduces  the spatiotemporal knowledge model proposed for representing meteorological events empirically described in newspaper articles. 
%Section \ref{sec:querying} 
%It also describes  module of LACLICHEV that assists data scientists in their expression of queries. The proposed queries can potentially explore the majority of entities (documents and event descriptions) talking about a particular type of climate event happening in a region during a time interval. 
Section \ref{sec:experimentation} describes the general architecture of LACLICHEV and the experiments we conducted for evaluating it. Section \ref{sec:relatedwork} studies approaches that promote datasets exploration for defining the type of analysis possible on top of them. Finally, \ref{sec:conclusion} concludes the paper underlying the contribution and discussing future work.

%¨¨¨¨¨¨¨¨¨¨¨¨¨¨¨¨¨¨¨¨¨¨¨¨¨¨¨¨¨¨¨¨¨¨¨¨¨¨¨¨¨¨¨¨¨¨¨¨¨¨¨¨¨¨¨¨¨¨
\section{Curating historical newspapers articles} \label{sec:architecture}
%¨¨¨¨¨¨¨¨¨¨¨¨¨¨¨¨¨¨¨¨¨¨¨¨¨¨¨¨¨¨¨¨¨¨¨¨¨¨¨¨¨¨¨¨¨¨¨¨¨¨¨¨¨¨¨¨¨¨

%Given digital newspapers collections and its associated terms frequency matrices, 

%\vspace{-0.8cm}

%\vspace{-1cm}
The objective of curating historical newspapers articles is to build a dataset of documents reporting meteorological events and associating them with metadata, providing as much information as possible about the reported event.
%Figure \ref{fig:process} shows 
The newspapers curation process is a semi-automatic process devoted to: 
%\begin{itemize}
%\item	
(i) find articles reporting climatologic events within newspapers articles collections; and
%\item	
(ii) geo-tag and store those articles that  report such events for building a database containing the climatologic event history. 
%
%\end{itemize}

% \begin{figure}[h]
% \begin{center}
% \includegraphics[width=0.90\textwidth]{figures/approach}
% \caption{Newspapers curation process}
% \label{fig:process}
% \end{center}
% \end{figure}
%

\begin{figure*}[h]
\begin{center}
\includegraphics[width=0.95\textwidth]{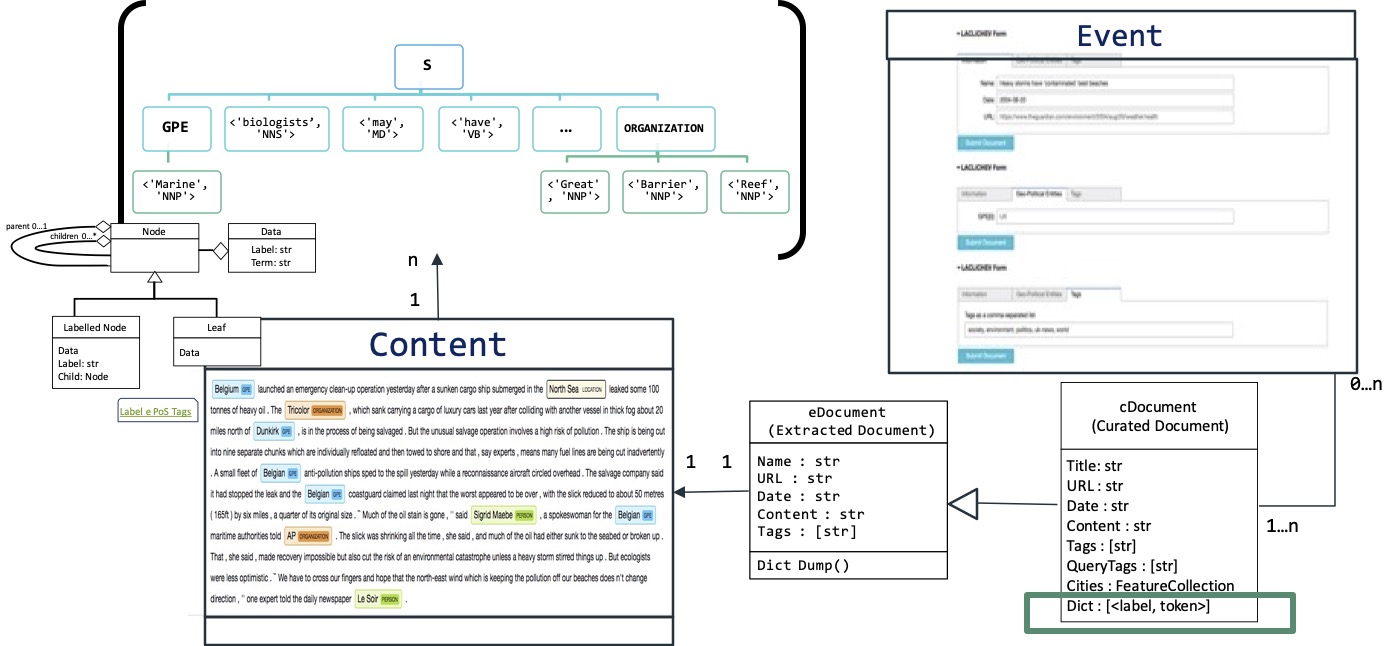}
\caption{Event Data Model}
\label{fig:curated-event}
\end{center}
\end{figure*}

%
%.  ...  ...  ...  ...  ...  ...  ...  ...  ...  ...  .. 
%\subsection{Newspapers Collections}
%.  ...  ...  ...  ...  ...  ...  ...  ...  ...  ...  .. 
 Curation tasks are performed on a collection of textual digital documents with minimum associated meta-data, particularly those used by digital libraries that own the collections. Each library adopts its meta-data schema, but in general, they specify the newspaper's name, the country, the day and number, the number of pages, the window time in which it circulated. Libraries export the meta-data schema used to describe these resources and align them to standards used by digital libraries. 
 %For example, the editions of the collection of Uruguayan newspapers published during the first 10 years of the XIX century.
%
%
The curation process generates data structures that provide an abstract representation of the content of each article describing an event. We propose a meteorological event knowledge model (see Figure \ref{fig:curated-event}) to represent climate event reports in digital documents. The objective is to describe events from different perspectives using the information from the articles and newspapers that report them and complete their description with domain knowledge also described in the model. As shown in Figure \ref{fig:curated-event} events are associated with the newspaper article(s) that describe them (reading from right to left). Each article can have meta-data that curates it, which points to its "raw" content that has been processed and annotated with linguistic labels.
Annotation and curation tasks can be recurrent and include a human-in-the-loop strategy for validating and adjusting results. For example, suppose an event is geo-tagged to associate it to a geographic location, and the event is described in an article about Montevideo news from Uruguayan newspapers collections. In that case, a human will verify that the geographic location refers to Montevideo in Uruguay and not in the US.

% Classes of documents associated with an event (class \textit{Curated Event} in the figure) containing variables that describe its characteristics like the date in which it happened or the geographical scope.

\vspace{-0.5cm}

The event knowledge model provides concepts for representing a \textit{meteorological event} according to the perspectives. Each aspect addressed by the knowledge model is implemented using different data structures with associated operations to support exploration actions. The following lines describe the different perspectives associated with an event and represented by the event model: descriptive, meteorologic, linguistic and knowledge domain profiles.
%Next we describe these data structures. These perspectives are described in the following lines.

%\begin{itemize}
    % \item  Profile. There is not a generic list of attributes because meteorological events are described in different way in historical newspapers articles, depending on the author. However, we can often collect information related to location, date, duration, scope, and damages.
    % \item  Linguistic perspective gathering the terms used for describing an event in one or several articles belonging to a given newspaper.
%     \item  Meteorological event profile perspective. Meteorological features (like millimeters of precipitations, wind speed, temperature, pressure, etc.) can be explicitly described in articles or they can be deduced according to the description of the event. For example, an event described happening in Montevideo and describing an overflow of the river, implies winds higher than 100 km/h., and rain of more than 10 mml/hour. This knowledge domain is used for completing the meteorologic features describing the reported event.
% \end{itemize}

%¨¨¨¨¨¨¨¨¨¨¨¨¨¨¨¨¨¨¨¨¨¨¨¨¨¨¨¨¨¨¨¨¨¨¨¨¨¨¨¨¨¨¨¨¨¨¨¨¨¨¨¨¨¨¨¨¨¨
\paragraph{Descriptive profile}
%¨¨¨¨¨¨¨¨¨¨¨¨¨¨¨¨¨¨¨¨¨¨¨¨¨¨¨¨¨¨¨¨¨¨¨¨¨¨¨¨¨¨¨¨¨¨¨¨¨¨¨¨¨¨¨¨¨¨
There is not a generic list of attributes used for describing a climatologic event in newspapers articles. Indeed, meteorological events are described in different ways in historical newspapers articles, depending on the author. However, we can often collect information related to location, date, duration, scope, and damages.
 Meteorological features (like millimetres of precipitations, wind speed, temperature, pressure, etc.) can be explicitly described in articles or deduced according to the description of the event. For example, an event reported in Montevideo and describing an overflow of the river implies winds higher than 100 km/h and rain of more than 10 ml/hour, according to the knowledge provided by meteorologists. This knowledge domain is used for completing the meteorologic features describing the reported event.
    
\begin{figure*}[h]
\begin{center}
\includegraphics[width=0.80\textwidth]{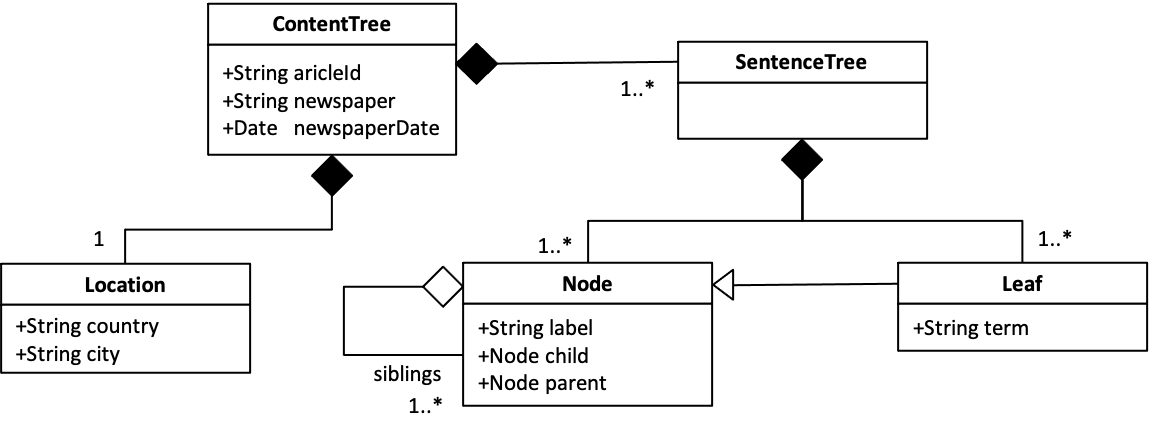}
\caption{UML class diagram representing the general structure of a content tree}
\label{fig:content-tree}
\end{center}
\end{figure*}

%¨¨¨¨¨¨¨¨¨¨¨¨¨¨¨¨¨¨¨¨¨¨¨¨¨¨¨¨¨¨¨¨¨¨¨¨¨¨¨¨¨¨¨¨¨¨¨¨¨¨¨¨¨¨¨¨¨¨
\paragraph{Linguistic perspective}
%¨¨¨¨¨¨¨¨¨¨¨¨¨¨¨¨¨¨¨¨¨¨¨¨¨¨¨¨¨¨¨¨¨¨¨¨¨¨¨¨¨¨¨¨¨¨¨¨¨¨¨¨¨¨¨¨¨¨
 gathers the terms used for describing an event in one or several articles belonging to a given newspaper. We propose a tree-based data structure, named {\em content tree} for representing the content of a historical newspaper article.
(see the UML class diagram in Figure \ref{fig:content-tree}). 
The tree corresponds to the grammatical analysis of each sentence in the textual content of the article commonly used in Natural Language Processing (NLP) techniques.
The {\bf content tree}, as shown below,  consists of a set of sentences. A {\bf sentence} is defined as a set of nodes representing grammatical elements of a sentence and leaves representing the terms composing a sentence in a specific article. 
 
% \begin{verbatim}
% - ContentTree: <articleId: String, 
%                 newspaper: String, newspaperDate: Date, 
%                 location: <country: String, city: String>, 
%                 sentences: {SentenceTree}>

% - SentenceTree: <nodes: {Node}, leafs: {Node}}>

% - Node: <label: Label, child: Node, parent: Node, 
%          siblings: {Node}>

% - Leaf: <term: String, parent: Node> extends Node

% \end{verbatim}

% In English, we use a grammatical model defined by the following Backus-Naur Form (BNF) specification:

% \begin{verbatim}
% <sentence> ::= <noun-sentence> | <verb-sentence>
% <noun-sentence> ::= <named-entity> <conjunction> <noun-sentence>
% <noun-sentence> ::= <noun>
% <verb-sentence> ::= <subject> <predicate>
% <subject> ::= <article> <noun>
% <predicate> ::= <verb> <direct-object>
% <direct-object> ::= <article> <noun>
% <article> ::= THE | A
% <noun> ::= "English nouns"
% <verb> ::= "English verbs"
% \end{verbatim}

\vspace{-0,5cm}

As shown  in Figure \ref{fig:content-tree},
a {\bf node} represents a type of grammatical element given in a specific linguistic model defined for a specific language. It is labelled adopting the entity labels produced by classic natural language processing tools known as Part Of Speech (POS) tags. For instance, \textit{noun, proper singular} (NNP), \textit{noun, plural} (NNS), \textit{verb, modal auxiliary} (MD), \textit{Geopolitical entity} (GPE), or Organization. In the case of subjects (NNP), they can be grouped into more general entities that identify geographic locations (GPE), places, names, and organization\footnote{A full list of POS tags can be found in \url{ https://www.cms.gov/}}. 
 
    A {\bf node} has a child, which is also a Node or a Leaf, and a set of siblings, which are other nodes. 
    A {\bf leaf} specializes a node, and it represents a term contained in the article. A term is a string and has a parent, a {\bf node} represents a POS tag.

%As shown in Figure \ref{fig:knowledge-model} 
Every article in a newspaper is associated with its content tree. Thereby, a data analyst or expert domain can explore the articles by navigating their content trees without reading the full content. For example, {\em retrieve articles reporting heavy storms in Uruguay in December 1810}.
% Nodes are related among each other through two relation types: instance, correlation. The relation of type correlation describes two terms that appear in the same sentence with a given distance given by the number of intermediate terms.

%¨¨¨¨¨¨¨¨¨¨¨¨¨¨¨¨¨¨¨¨¨¨¨¨¨¨¨¨¨¨¨¨¨¨¨¨¨¨¨¨¨¨¨¨¨¨¨¨¨¨¨¨¨¨¨¨¨¨
\paragraph{Meteorological   perspective}
%¨¨¨¨¨¨¨¨¨¨¨¨¨¨¨¨¨¨¨¨¨¨¨¨¨¨¨¨¨¨¨¨¨¨¨¨¨¨¨¨¨¨¨¨¨¨¨¨¨¨¨¨¨¨¨¨¨¨
characterizes the event with attributes used to describe it in one or several newspaper articles. Nevertheless, not all the attributes can necessarily have a value since there might not be any evidence within the articles that report it. Attributes, like the date of the event, its geographical scope, or the location of the damaged regions, are computed by navigating through the {\em content tree} of every article reporting the event.
\begin{verbatim}
ClimateEvent: <date: Data, duration: <init:Date, end:Date> 
        scope:{<locationName: String, long: Float, lat: Float>}
        name: String, damages: {String}>
\end{verbatim}

%¨¨¨¨¨¨¨¨¨¨¨¨¨¨¨¨¨¨¨¨¨¨¨¨¨¨¨¨¨¨¨¨¨¨¨¨¨¨¨¨¨¨¨¨¨¨¨¨¨¨¨¨¨¨¨¨¨¨
\paragraph{Knowledge domain perspective}
%¨¨¨¨¨¨¨¨¨¨¨¨¨¨¨¨¨¨¨¨¨¨¨¨¨¨¨¨¨¨¨¨¨¨¨¨¨¨¨¨¨¨¨¨¨¨¨¨¨¨¨¨¨¨¨¨¨¨
 describes meteorological events using knowledge domain statements created by experts of the National Library of Uruguay. This knowledge has been associated with events through manual analysis of newspapers collections and interacting with meteorologists. This knowledge can help interpret the empirical information reported in the articles and complete the information associated with the event description. For example, if the river was flooding due to a storm, it is possible to estimate the wind speed and the approximate litters of rain. The knowledge domain perspective is modelled as a glossary. 
 
 %Figure \ref{fig:glossary} shows the intuition of its structure.

%\textbf{Cual es la procedencia de este conocimiento? Deberiamos explicar su contenido?????}

% \begin{figure*}[h]
% \begin{center}
% \includegraphics[width=0.90\textwidth]{figures/glossary.pdf}
% \caption{Climate events glossary}
% \label{fig:glossary}
% \end{center}
% \end{figure*}
% \vspace{-0,9cm}

%.  ...  ...  ...  ...  ...  ...  ...  ...  ...  ...  .. 
\subsection{Building a climatologic event history}
%.  ...  ...  ...  ...  ...  ...  ...  ...  ...  ...  .. 
Given a documents' collection and its associated data structures describing the content of its articles, the data scientist can explore articles to determine whether they report climatologic events. This phase integrates the human-in-the- loop. The reason is that newspaper articles' use of colloquial terms can be tricky and refer to metaphors that might not denote a climatologic event. Language subtilities are not easy to handle manually, mainly because we are dealing with a language used some centuries ago, which increases the challenge of classifying the content of the articles. During this phase, articles referring to climatologic events are  
 geo-temporally tagged to associate them with  the region and/or time window in which they happened. Tags are validated by the data analyst. Since the result can contain a significant number of articles, the user can use three tools for understanding the content of the result. The tools let her manipulate a terms frequency matrix, a terms heat map. She can also explore the content of the article text using a view that provides information about the context in which the terms potentially describing an event appear in the document. For example, the name of geographic locations in the document might refer to the location of the event and the region that it touched, and a list of geopolitical entities (e.g., school, public building, etc.) to determine the damages caused by the event.

%Likewise, 
The data analyst can perform the following actions: \\
%\begin{itemize}
%\item	
\noindent
- Correct the terms associated with climatologic events that might not be used in such a sense in the text. Indeed, some social and political demonstrations are often described as climatologic events. For a classic automatic text analysis process, this can be not easy to identify and filter. For example, an article entitled {\em Stormy weather within the ails of the senate in Ecuador} has nothing to do with a climatologic event but with a political one.

%\item
\noindent
- Determine whether personal names correspond to the event's name (e.g. hurricane or storm name). If that is the case, this information will be used for inserting the event in the history.

%\item	
\noindent
- Verify whether the names of cities, regions and countries correspond to geographic entities. The system underlines the names of patronyms, and the data analyst can see the location of the possible geographic entities. Thus, the user can also confirm whether the article refers to the geographic zone that she is searching for. For instance, if "Santa Clara" is underlined, it can refer to a place, city, or village.

%\item	
\noindent
- Determine the date of the event and its characteristics. The temporal terms and adjectives are also underlined to let the data analyst click on those that describe the event.

%\item
\noindent
- Determine the type of damages caused by the event by exploring those terms that describe such kind of information.
 %\end{itemize}

\begin{figure*}[h]
\begin{center}
\includegraphics[width=0.95\textwidth]{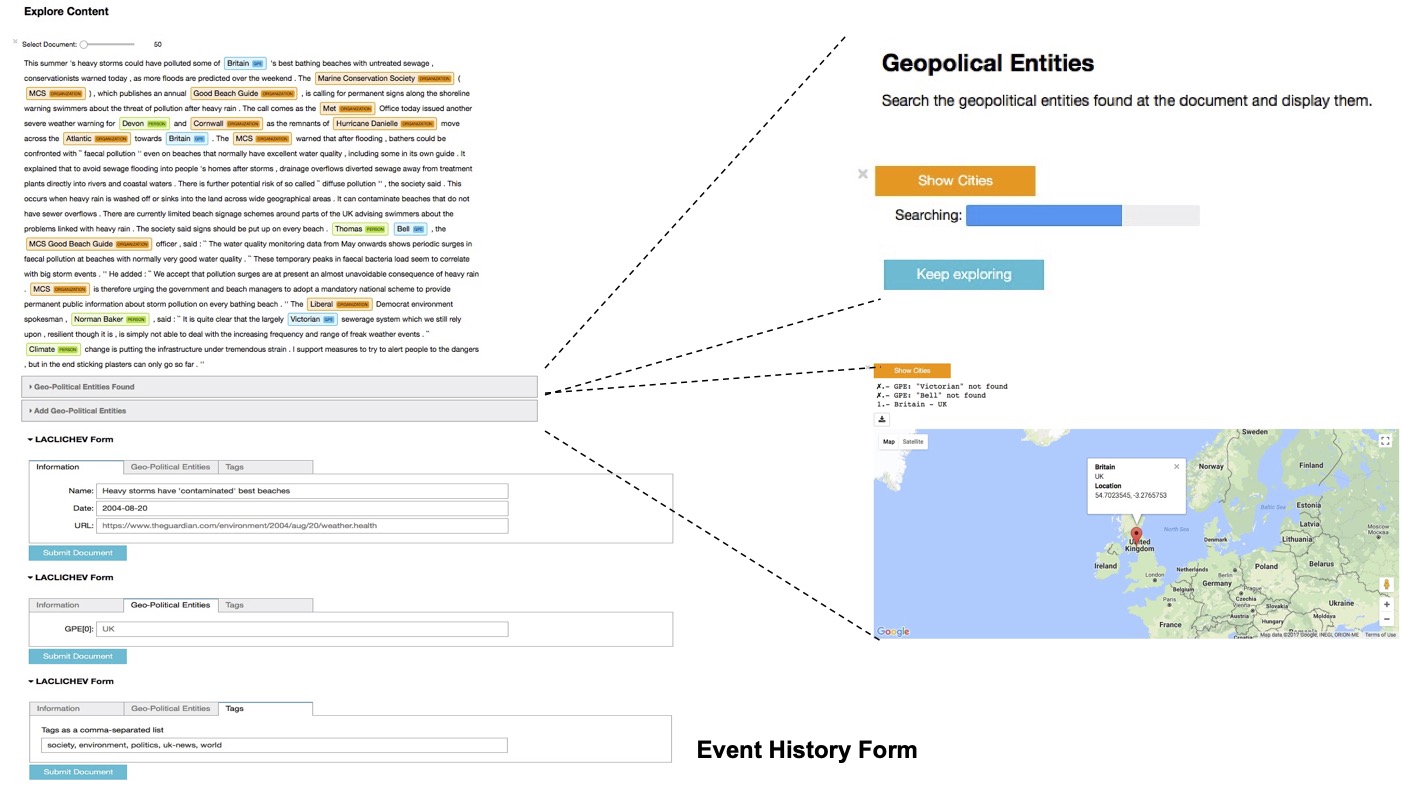}
\caption{Event curation process interface for tagging events}
\label{fig:event-tagging}
\end{center}
\end{figure*}

The previous actions are used to adjust the representation of the articles' content and identify meteorological events more accurately since the data analyst or domain expert knowledge is used (see Figure \ref{fig:event-tagging} showing LACLICHEV interfaces for curation). 
%Once the events have been validated by experts, they can be stored in a knowledge events history.

%\vspace{-0.65cm}

%.  ...  ...  ...  ...  ...  ...  ...  ...  ...  ...  .. 
\subsection{Climatologic Events History}
%.  ...  ...  ...  ...  ...  ...  ...  ...  ...  ...  .. 
%The previous actions are used to adjust the representation of the articles' content and identify climatologic events and make them part of the history of climatologic events. 
Once the events have been validated, the data scientist can use a form to define and store the event. 
Meta-data is stored in persistence support, a key-value or a document store depending on the technology adopted by each library, whereas the raw documents remain archived in a different server or the same store. 
%
%The approach uses a document store (i.e., MongoDB) for storing geo-temporally tagged climatologic events. 
The climatologic event's history provides an interface for performing querying and analytics tasks on top of it. For example, locate events that happened during the XIX century, enumerate and locate the most famous climatological events in the region, create a heat map of the events in Latin America that happened in the last ten years of the XIX century.

% The digital collection can be initially queried by filtering the documents by region or country or by year. Digital libraries offer front ends for performing this classic information retrieval process. For example, select newspapers published in Uruguay (i.e., geographic filter)  between 1800-1810 (i.e., temporal filter).

%.  ...  ...  ...  ...  ...  ...  ...  ...  ...  ...  .. 
\subsection{Exploring the collections of digital newspapers}
%.  ...  ...  ...  ...  ...  ...  ...  ...  ...  ...  .. 
%  %The previous actions are used to adjust the representation of the articles' content and identify climatologic events and make them part of the history of climatologic events. 
% Once the events have been validated, the data scientist can use a form to define and store the event. 
% Meta-data is stored in persistence support, a key-value or a document store depending on the technology adopted by each library, whereas the raw documents remain archived in a different server or the same store. 
% %
% %The approach uses a document store (i.e., MongoDB) for storing geo-temporally tagged climatologic events. 
% The climatologic event's history provides an interface for performing querying and analytics tasks on top of it. For example, locate events that happened during the XIX century, enumerate and locate the most famous climatological events in the region, create a heat map of the events in Latin America that happened in the last ten years of the XIX century.

 %Meteorological events exploration through digital historical newspapers data collections is done on top of the climatologic event's history. 
 %\vspace{-0.62cm}
 Newspaper articles are explored by conjunctive or disjunctive keyword queries, where keywords can belong to several vocabularies. For example, search articles reporting heavy storms and rivers flooding. The query expressed by a data analyst is automatically completed by using rewriting techniques that consider synonyms, more specific or more general concepts \cite{10.1007/978-3-319-44066-8_9}. Thus,  three tools can be used for exploring meteorological events depending on expert knowledge of what she/he is looking for:
 
%  The rewriting process produces several proposals that the data analyst can adjust and then choose to be evaluated.
% Each chosen query is evaluated using information retrieval techniques, including the article's text stemming for extracting the terms, constructing a frequency matrix that provides occurrence statistics of the representative terms of the text content within a collection of documents. 

% In general, information retrieval processes do not exhibit this matrix; it is an internal data structure representing the content of the documents and is used to answer queries. In our approach, this frequency matrix is accessible to the data scientist because it provides an aggregated view of the content of a documents collection. In our approach, we also compute and exhibit a terms heat map for a given documents collection to provide a more economical (i.e., consolidated) view of the collection's content. Our approach provides an interactive interface that let data scientist manipulated these data structures for defining the piece of collections she wants to explore.

% The data scientist can explore them and then decide whether the collection can describe climatologic events and the documents that might be closer to her requirements. She can decide eventually to explore some documents directly or reformulate the query. Once a result containing articles that potentially answer the query has been computed, the user can explore the result and validate the selection elements during the next step of the data exploration workflow. 

%\begin{itemize}
 %   \item 
\noindent    
    \textit{- Filtering}. Retrieving factual information. For example, filtering events by region or country or by year. For example, Uruguay for the country and between 1800–1810 for the temporal filter.
   
    %\item 
    \noindent  
    \textit{- Term frequency}. Understanding the content of digital newspaper collection through the vocabulary used in its articles. Therefore, LACLICHEV exposes the terms frequency matrix and a terms heat map under an interactive interface. The domain expert can see which are, statistically, the terms most used in the articles, group documents according to the terms used, choose articles using a specific term.
    
    %\item 
    \noindent  
    \textit{- Additional information}. Exploring the content of a specific article using a view that provides information about the name of geographic locations in the document that might refer to the location of the event and the region that it touched, and a list of geospatial features (e.g., school, public building, etc.) to determine, for example, the damages caused by the event. 
%\end{itemize}

%----------------------------------------------------------------------------
\section{LACLICHEV in action}\label{sec:experimentation}
\begin{figure*}[h]
\begin{center}
\includegraphics[width=0.90\textwidth]{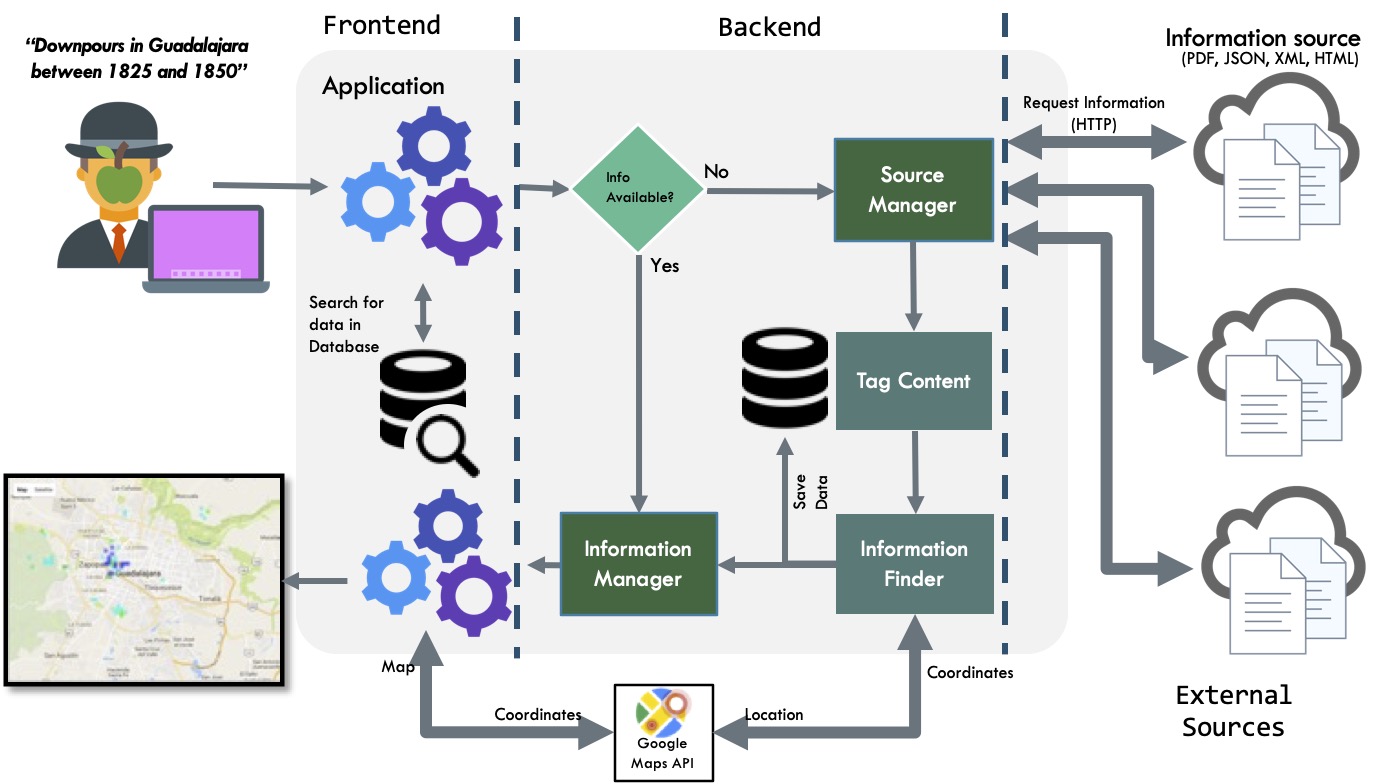}
\caption{General functional architecture of LACLICHEV}
\label{fig:architecture}
\end{center}
\end{figure*}

%----------------------------------------------------------------------------
We have implemented LACLICHEV, a data collection exploration environment accessible as a Jupyter notebook. It is a client-server system that uses the platform Jupyter for executing the human-in-the-loop tasks that implement the data exploration process. Figure  \ref{fig:architecture} shows the general architecture of LACLICHEV organised into three layers: (i) frontend giving access to the event history containing curated articles reporting climatologic events and providing tools for curating articles and creating events descriptions; (ii) backend with the climatologic event history and tools for pre-processing newspaper articles; (iii) external layer connecting to documents providers that are available through servers accessible in the web and through APIs exported by libraries.
%composed of five modules: (i) curation, (ii) exploration, (iii) knowledge base, (iv) query processing, and (v) spatio-temporal component.

%\begin{itemize}
%    \item find articles reporting climate events within digital collections available in existing digital libraries repositories;
%    \item geo-tag interactively and store those articles that really report such events for building a climate event database. 
%\end{itemize}

%\vspace{-0.9cm}
%Therefore LACLICHEV relies on a knowledge graph the integrates a thesaurus classifying climate event types, Wordnet and a glossary defining meteorologic characteristics of climate events.

% Next, we briefly  describe  these  components,  illustrating  the  descriptions  with  some examples.

%-------------------------------------------------------
\subsection{Building a Latin American climatologic event history}
%-------------------------------------------------------
We have worked with the national libraries of Mexico, Colombia, Ecuador and Uruguay to access their newspapers digital collections. 
For our experiments, we worked with the collections of the XVIII and XIX centuries of newspapers written in Spanish with the linguistic variations of three countries Mexico, Colombia, Ecuador and Uruguay. 

%We used ontologies of meteorological terms created by analyzing the collections that we deal with this work. 

%¨¨¨¨¨¨¨¨¨¨¨¨¨¨¨¨¨¨¨¨¨¨¨¨¨¨¨¨¨¨¨¨¨¨¨¨¨¨¨¨¨¨¨¨¨¨¨¨¨¨¨¨¨¨¨¨¨¨
%\subsection{Pre-processing newspapers collections}
%¨¨¨¨¨¨¨¨¨¨¨¨¨¨¨¨¨¨¨¨¨¨¨¨¨¨¨¨¨¨¨¨¨¨¨¨¨¨¨¨¨¨¨¨¨¨¨¨¨¨¨¨¨¨¨¨¨¨

\begin{figure}[h]
\begin{center}
\includegraphics[width=0.99\textwidth]{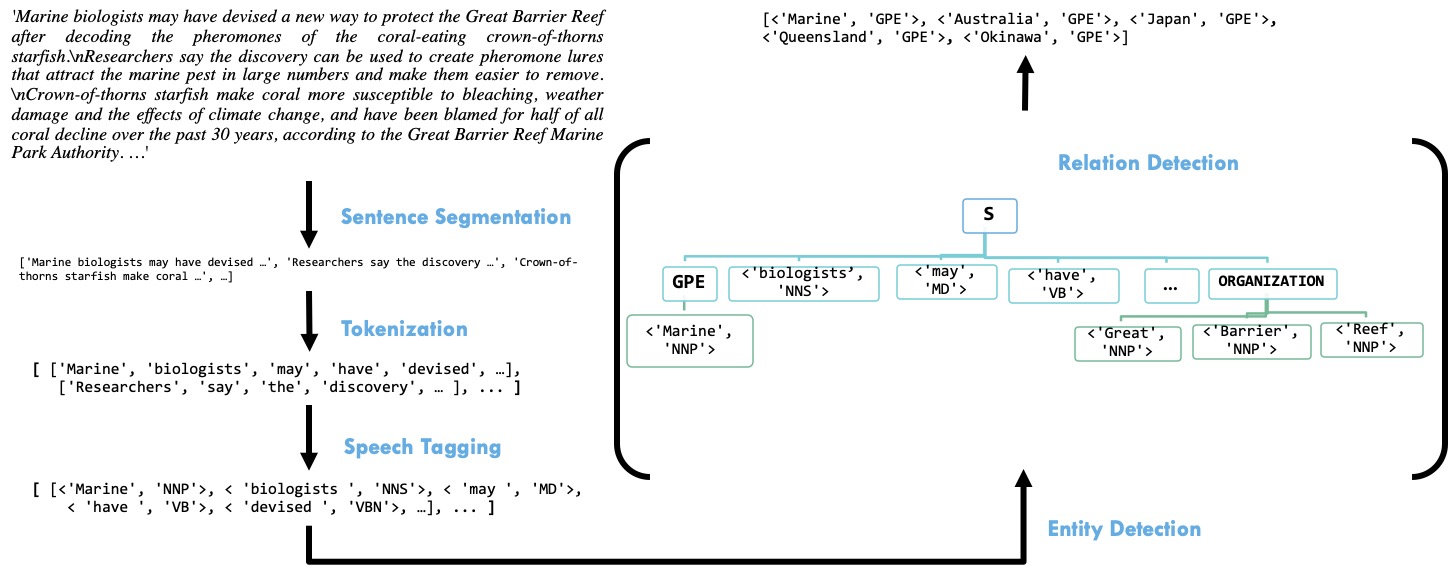}
\caption{Pre-processing text pipeline}
\label{fig:preprocessing-pipeline}
\end{center}
\end{figure}

We curated collections and generated the vocabulary used on articles identified as reporting a meteorological event.
%(see Figure \ref{fig:preprocessing-pipeline}). 
Digital newspaper collections remain in the initial repositories that belong to the libraries. Then, terms and links to the OCR (Optical Character Recognition) archives containing documents with articles reporting meteorological events were stored in distributed histories managed in each country.  As shown in  Figure \ref{fig:preprocessing-pipeline}, the process consists of five steps usually used in natural language processing techniques: sentence segmentation, tokenization, speech tagging, entity and relation detection.  LACLICHEV implements these phases on Python, relying on the NLTK library.

% The first phase of the pre-processing process of newspapers leads to graphs representing the content of the articles and classic inverse index and frequency matrices used for performing exploration queries.

\begin{figure*}[h]
\begin{center}
\includegraphics[width=0.95\textwidth]{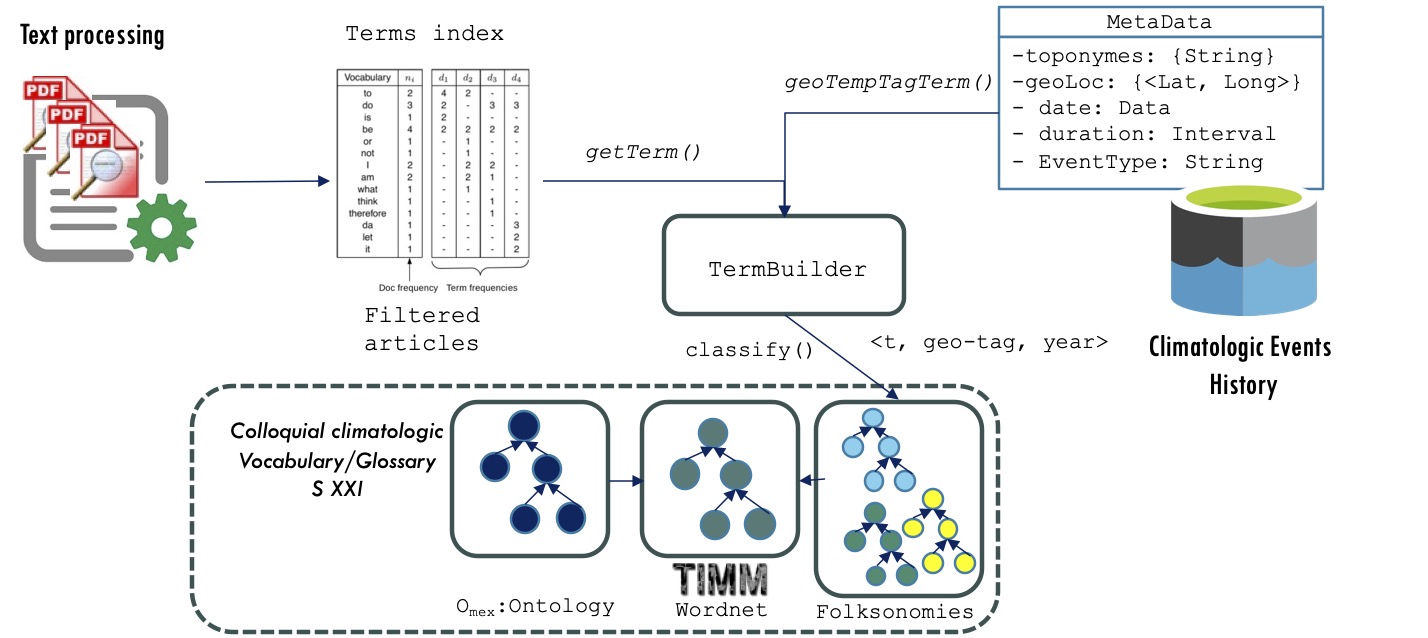}
\caption{Collecting colloquial vocabulary}
\label{fig:colloquial}
\end{center}
\end{figure*}

%\vspace{-0,5cm}
Besides curating the data collections content, we wanted to discover linguistic variations in different Latin American countries to describe meteorological events. People's language and variations can picture civilians' perception of these events, consequences, and associated explanations. Thus, local vocabularies were created out of the terms used in newspapers' articles (see Figure \ref{fig:colloquial}). For example, referring to a storm as a stormtrooper\footnote{In Mexico a storm is called a "chaparrón" and in Uruguay it is called a "chubasco".} Then we updated and enriched through queries, exploration and analytic activities these vocabularies through human-in-the-loop actions. Data analysts tagged "colloquial" terms used to describe climatologic events and associated them with more scientific terms. These terms can be then used for defining keyword queries for exploring newspapers datasets.

%\vspace{-0,5cm}

%¨¨¨¨¨¨¨¨¨¨¨¨¨¨¨¨¨¨¨¨¨¨¨¨¨¨¨¨¨¨¨¨¨¨¨¨¨¨¨¨¨¨¨¨¨¨¨¨¨¨¨¨¨¨¨¨¨¨
\subsection{Querying, exploring and curating data collections}
%¨¨¨¨¨¨¨¨¨¨¨¨¨¨¨¨¨¨¨¨¨¨¨¨¨¨¨¨¨¨¨¨¨¨¨¨¨¨¨¨¨¨¨¨¨¨¨¨¨¨¨¨¨¨¨¨¨¨

LACLICHEV proposes a frontend for data scientists to: 

%\begin{itemize}
    %\item 
    \noindent
    - Query the event's history of already tagged events. The queries can be keyword oriented (e.g. locate the most famous events in Mexico during the XVIII century).
    Users decide to use some terms that can belong to any of the vocabularies generated in the pre-processing phase. LACLICHEV applies then query rewriting techniques to extended user expressed queries with synonyms, subsuming and general terms. The particular characteristic of this task is that the user (i.e., data analyst) can interact and guide the process according to her/his knowledge and expectations about what she/he expects to explore and search. %This user guide is called human in the loop.
    
    %\item  
     \noindent
    - Perform analytics operations and analyse results are generally presented within maps (e.g. how did rainy periods evolved in the region?). In the current version of LACLICHEV analytics queries cannot be expressed in the frontend. They are implemented manually through notebooks running on top of the event history. 

%\item 
 \noindent
- Exploring data collections for curating them and building the history of climate events.

%\item 
 \noindent
-  Manage vocabularies, adding terms, guiding their classification and studying the linguistic connections between the terms used in the different countries.

%\end{itemize}

Data analysts can express queries that can potentially explore historical newspapers content to find articles describing meteorological events. The aim is to have a good balance between precision and recall despite the ambiguity of the language (Spanish variations in naming climate events). The domain experts must express "clever" queries that can exploit the collections to achieve this goal. 

Queries can be initial conjunctive and disjunctive expression combining terms chosen from the built-in vocabularies or not.
Then, queries are rewritten in an expression tree where nodes are conjunction and disjunction operators and leaves are terms, according to 
an input query expressed as a conjunction and disjunction of terms potentially belonging to a meteorological vocabulary.
Our approach for rewriting queries is based on a "queries as an answer" process.
This technique rewrites user queries into queries that can produce more precise results according to the explored dataset content.
%gives a list of queries based on the information within a given data collection. The key challenge is identifying interesting queries to return. 
Queries as answers proposed by LACLICHEV consists of a list of frequently used queries. Thus, we focus on the following aspects:

%¨¨¨¨¨¨¨¨¨¨¨¨¨¨¨¨¨¨¨¨¨¨¨¨¨¨¨¨¨¨¨¨¨¨¨¨¨¨¨¨¨¨¨¨¨¨¨¨¨¨¨¨¨¨¨¨¨¨
\paragraph*{Extending query alternatives using hypernyms and synonymes}
%¨¨¨¨¨¨¨¨¨¨¨¨¨¨¨¨¨¨¨¨¨¨¨¨¨¨¨¨¨¨¨¨¨¨¨¨¨¨¨¨¨¨¨¨¨¨¨¨¨¨¨¨¨¨¨¨¨¨
Given an initial conjunctive/disjunctive  query is rewritten by extending it with general and more specific terms, synonyms, etc. The terms used to express the query belong to colloquial vocabulary for denoting meteorological events. The rewriting process can be automatic or interactive, in which case the system proposes alternatives and the user can validate the proposed terms. For example, if the query is {\em "heavy storms"}, the query can be completed by adding {\em "heavy stormtrooper"}, {\em "heavy storm dust"}.

%\textbf{Cual es la procedencia del colloquial vocabulary????}

We use Wordnet\footnote{http://timm.ujaen.es/recursos/spanish-wordnet-3-0/} for looking for associated terms and synonyms that help address concepts used in different Spanish speaking countries. We do not translate the query terms to other languages because our digital data collections consist of newspapers written in Spanish.
 LACLICHEV allows equivalent terms searching to morph a query. For a new term, LACLICHEV generates a node with the operator {\sf\small and} then connects the initial term with the equivalent terms in a disjunctive expression subtree. Thereby, more general terms are collected and connected to the initial term with these terms in a conjunctive expression subtree. The result is a new expression tree corresponding to an extended query {\sf\small Q$_{ExT}$}. The query morphing algorithm behind LACLICHEV is described in \cite{vargas2021towards}.

%¨¨¨¨¨¨¨¨¨¨¨¨¨¨¨¨¨¨¨¨¨¨¨¨¨¨¨¨¨¨¨¨¨¨¨¨¨¨¨¨¨¨¨¨¨¨¨¨¨¨¨¨¨¨¨¨¨¨
\paragraph*{Extending query alternatives using cultural terms}
%¨¨¨¨¨¨¨¨¨¨¨¨¨¨¨¨¨¨¨¨¨¨¨¨¨¨¨¨¨¨¨¨¨¨¨¨¨¨¨¨¨¨¨¨¨¨¨¨¨¨¨¨¨¨¨¨¨¨
Use local vocabularies for generating new query expression trees that substitute the terms used in {\sf\small Q'$_{ExTi}$} with equivalent terms used in a target country (e.g., blizzard instead of a heavy storm). This will result in  transformed expression trees each one using the terms of a country ({\sf\small Q''$_{ExT1}$} ... {\sf\small Q''$_{ExTj}$}) \cite{vargas2018computing}.

A series of what we call "folksonomies" in a metaphorical way, which are vocabularies created through the processing of the vocabulary of newspaper articles. We create and feed each vocabulary according to the country of origin of the processed newspaper article. This lets us extract the vocabulary used during the XVIII and XIX centuries for describing meteorological events in different Latin American countries  (i.e. Mexico, Colombia, Ecuador, and Uruguay). Using this information LACLICHEV can answer the following queries: 
{\em How have terms used to describe climatological events changed between  XIX–XX c.? Which are standard terms used to describe climatological events across Latin American countries? Which is the distance between terms used in  XIX–XX c.?
Which are the most popular terms used in  XIX c. for describing climatological events? }

%¨¨¨¨¨¨¨¨¨¨¨¨¨¨¨¨¨¨¨¨¨¨¨¨¨¨¨¨¨¨¨¨¨¨¨¨¨¨¨¨¨¨¨¨¨¨¨¨¨¨¨¨¨¨¨¨¨¨
\paragraph*{Defining filters using knowledge domain}
%¨¨¨¨¨¨¨¨¨¨¨¨¨¨¨¨¨¨¨¨¨¨¨¨¨¨¨¨¨¨¨¨¨¨¨¨¨¨¨¨¨¨¨¨¨¨¨¨¨¨¨¨¨¨¨¨¨¨
We also use domain knowledge for rewriting the queries. We have a knowledge base provided by domain experts that contains some meteorological event rules. For example, rules state that in the presence of a heavy storm:
%\begin{itemize}
%\item 
R1. the wind speed is higher that 118 km/hr;
%\item 
R2.  the rivers can grow and produce big waves;  
%\item 
R3. there are rains between 2,5 7,5 mm/hr;
%\item 
R4. the range of surface that can be reached by a 100 km wind speed storm is of 1000 km.
%\end{itemize}

Our approach uses this information for generating possible queries that help the domain expert better precise her/his query or define several queries that can be representative of what she/he is looking for. For example, the previous initial query "Q$_{1}$: heavy storm" is rewritten into new additional queries: 
% \begin{itemize}
 %\item 
 "Q$_{11}$: heavy storm {\em or storm with  wind speed $>$ 100 km}" (using R1). 
 % \item 
  "Q$_{12}$: storms with 100 km speed that reached Mexico City" (using R2 and knowing the initial point and geographic information). 
  %\item 
  "Q$_{13}$: storms touching villages 500 km around Mexico city happening in the same period" (R4). 
%\end{itemize}
%
Instead of having a long query expression, our approach proposes sets of queries that the domain expert can choose and combine.

\paragraph*{Analytics queries}
%¨¨¨¨¨¨¨¨¨¨¨¨¨¨¨¨¨¨¨¨¨¨¨¨¨¨¨¨¨¨¨¨¨¨¨¨¨¨¨¨¨¨¨¨¨¨¨¨¨¨¨¨¨¨¨¨¨¨
The climatological event's history provided and maintained by LACLICHEV can visualize information and perform analytical tasks. For example, LACLICHEV can answer spatio-temporal queries like: {\em Q$_1$ Locate climatologic events in the XVII c.}, {\em Q$_2$ Enumerate \& locate the most famous events in the region } and {\em Q$_3$ Create a heat map of events in LATAM in the last years of the XIX c.} The objective is to answer analytics queries that imply aggregating information stored in the event's history. For example, {\em How did rainy time evolve in time in the region?}, {\em In which way was climate different between XVII and XIX c.?
How did vocabulary evolve from colloquial to scientific and standardized in the XX c.? 
} In future versions, LACLICHEV is willing to answer prediction queries like {\em Could it have been possible to predict the evolution of climate behaviour from the data in XVIII and XIX c.?} This type of queries requires the collection, curation and preparation of more newspapers articles and other complementary data. This concerns future work.

%----------------------------------------------------------------------------
\section{Related work}\label{sec:relatedwork}
%----------------------------------------------------------------------------

% On top of that, humans are relatively poor at interactions with massive data, especially if they are unstructured; according to Carole Goble “from a big data perspective, the challenges are around finding the slices, views or ways into the dataset that enables you to find the bits that need to be edited, changed”. 

The emergence of the notion of data exploration provides different perspectives of the data and tools for helping data scientists choose and compound datasets adapted for target experiments \cite{kersten2011researcher}. 
%The tools include functions like “data grooming”, which denotes the process of transforming raw data into analysable data with various data structures. 
%Other approaches focus on transforming human-readable data into machine-readable data considering inconsistencies in data formatting given that they are produced under different conditions. The idea is to exhibit processes, digital spaces, and systems that host datasets and give them access to understand how conditions data can be processed.

% To characterize data exploration, we need to compare it to data querying and associate it to visualization that provides techniques to understand the content of collections in an aggregated manner. 
% Data querying aims to obtain all the data tuples respecting a defined often in the objective of answering a related question with correct and complete results. This means knowing what data is in the database and what structure. In contrast, in raw digital data collections, this cannot be guaranteed. 
Despite the availability of datasets, very often, users are unsure which patterns they want to find. Data exploration \cite{kersten2011researcher} addresses this requirement and proposes strategies for going into the whole or samples of datasets to understand their content for determining the type of questions they can answer. 
Some of the techniques used for fulfilling this particular requirement are data grooming, multi-scale queries, result set post-processing, query morphing, and queries as answers. 
Data grooming denotes the process of transforming raw data into analysable data with various data structures.
Multi-scale queries propose to split a query into multiple queries executed on different fragments of the database and then perform a union of those queries. %This allows scaling the size of the query as the user gets more confident in her query. 
%Result set post-processing and query morphing go on the premise that the user probably does not need the exact answer to a query. 
Result set post-processing assumes an array of simple statistical information such as min, max, and mean to be more helpful, especially on massive data sets. 
%Query morphing assumes queries can be missformulated. Query morphing still focuses on answering the query given by the user but will also use a small portion of resources in searching data around the original query. 
%
%Another trend regarding data exploration is to tackle the lack of knowledge a user may have on the dataset. 
Query morphing and queries as an answer are rewriting techniques that compute alternative queries (e.g. adding terms) that can potentially better explore a dataset than an initial query. Approaches such as interactive query expansion (IQE)  \cite{ruthven2003re,belkin2008some,goswami2017exploring} have shown the importance of data consumers in the data exploration process. Users’ intention helps to navigate through the unknown data, formulate queries and find the desired information.  In most of the occurrences, user feedback acts as vital relevance criteria for next query search iteration. 

Existing solutions are not delivered in integrated environments that data analysts can comfortably use to explore data collections. 
%and design analytics settings. 
The technical effort is still necessary to combine several tools to explore and process datasets and go from raw independent data sets to knowledge, %understanding and prediction, 
for example, on climate change. 
% Therefore, our research aims to tailor a data exploration environment that can help explore digital data collections using a human in the loop approach.
%  Existing solutions are not delivered in an integrated manner in environments that data analysts can comfortably use to explore data collections and design analytics settings. Technical effort is still necessary to combine several tools to explore and process data sets and go from raw independent data sets to knowledge, understanding and prediction, for example, on climate change. 
Therefore, LACLICHEV aimed to tailor a data exploration environment that could help explore digital datasets using a human-in-the-loop approach.

%----------------------------------------------------------------------------
\section{Conclusion and Future Work}\label{sec:conclusion}
%----------------------------------------------------------------------------
This paper presented our approach for exploring newspaper digital collections for building knowledge about the history of climate change in Latin America. Using well-known information retrieval and analytics techniques, we developed a data exploration environment that provides tools for understanding the content of collections. Rather than directly querying collections for searching documents or performing data analytics operations (statistics, correlations), the objective is to let data scientists understand the content of the collections and then decide what kind of queries to ask. We used digital newspapers collections for applying such techniques for building and analyzing the history of climate change in Mexico, Colombia, Ecuador and Uruguay. The work reported here is the first step towards this ambitious challenge. We continue enriching data collections, developing and testing solutions for generating and sharing step by step this history.

% %----------------------------------------------------------------------------
% \section*{Acknowledgements}
% %----------------------------------------------------------------------------
% We thank the master student Santiago Ruiz Angulo of the Universidad Aut\'onoma de Guadalajara who implemented the first version of LACLICHEV during his internship at the Barcelona Super Computer Centre, Spain funded by the CONACYT “beca mixta” fellowship program of the Mexican government.

% %----------------------------------------------------------------------------
% \section*{Funding}
% %----------------------------------------------------------------------------
% This work has been supported by the Pan-American Institute of Geography and History (PAIGH). 

 \bibliographystyle{splncs04}
 \bibliography{acmart}

\end{document}